\documentclass[10pt]{article} 
\usepackage[utf8]{inputenc}
\usepackage{authblk}
\usepackage{tikz} 
\usepackage{tabularx}
\usepackage{appendix}

\usepackage[utf8]{inputenc}
\usepackage{appendix}

\usepackage[utf8]{inputenc}
\usepackage{authblk}
\usepackage{tikz} 
\usepackage{tabularx}
\usepackage{appendix}
\usepackage[english]{babel}
\usepackage{amsthm}
\usepackage{lineno}

\usepackage{shorthand}
\usepackage{amsmath}

\usepackage{biblatex} 
\bibliography{bibliography.bib}

\title{\textbf{ 
A Simple Linear Algebraic Approach to Capture the Dynamics of the Circular Flow of Income}}

\author{by\\ Aziz Guergachi \ and \  Javid Hakim}

\begin{document}
\maketitle
\begin{abstract}
    \noindent This article has one single purpose: introduce a new and simple, yet highly insightful approach to capture, fully and quantitatively, the dynamics of the circular flow of income in economies. The proposed approach relies mostly on basic linear algebraic concepts and has deep implications for the disciplines of economics, physics and econophysics.   \\  \\   \\  
\end{abstract}

\noindent \textbf{\Large The act of selling and buying as \textit{a} driver of money circulation:}
\\

\noindent A fundamental concept that underlies the notion of scarcity, which is the central core of economics (e.g.,\cite{mankiw}, page 4; \cite{samuelson}, page 4), is the fact that there are buyers who desire to acquire the scarce goods and services in the market \cite{PengAziz}. If no one is interested in these goods and services, then they would be irrelevant to the study of economics, even if they were scarce. The approach we propose in this paper to describe the circular flow of income focuses on the very exchange of goods and services between sellers and buyers, and attempts to capture the dynamics of this exchange from the ground up.\\

\noindent To describe the basic idea of this approach, let us imagine a community of three economic agents, who sell goods and services to each other in exchange for money, all in a closed economy. We refer to these three agents as A, B and C, also known as Alice, Bob and Carol (to reuse the fictional characters introduced by cryptographers Rivest et al. \cite{Rivest}). As a result of the business exchange taking place among them, Alice, Bob and Carol earn income. Let us denote the incomes earned by Alice, Bob and Carol up until time $t$ and expressed in monetary units as $x_1(t)$, $x_2(t)$ and $x_3(t)$ respectively. Between time $t$ and time $t+1$, Alice will sell goods and services to Bob who will then pay Alice a certain price $p$ for these goods and services. This price $p$ paid by Bob to Alice will come out of Bob's income $x_2(t)$ at time $t$; it will be a fraction of $x_2(t)$. Let us denote this fraction of $x_2(t)$ paid by Bob to Alice during the time interval $[t,t+1]$ as $f_{12}(t)$. In a similar way, Alice will also sell goods and services to Carol between $t$ and $t+1$ and will, as a result, earn an additional income of $f_{13}(t) \times x_3(t)$, where $f_{13}(t)$ is the fraction of $x_3(t)$ paid by Carol to Alice between $t$ and $t+1$ in exchange of the purchased goods and services. Thus, the total income earned by Alice over the time interval $[t,t+1]$ as a result of her doing business with other economic agents is:
\begin{equation}
    x_1(t+1) = f_{12}(t) \, x_2(t) + f_{13}(t) \, x_3(t) 
\end{equation}
\\
Let us now focus on Bob as a seller of some different type of goods and services to other economic agents, namely Alice and Carol who would then have to pay some fractions $f_{21}(t)$ and $f_{23}(t)$ of their respective incomes $x_1(t)$ and $x_3(t)$ at time $t$ to Bob. Thus, over the time interval $[t,t+1]$, Bob earns a total income of:

\begin{equation}
    x_2(t+1) = f_{21}(t) \, x_1(t) + f_{23}(t) \, x_3(t) 
\end{equation}
as result of him conducting business with the other agents.\\~\\

\noindent In a similar way, Carol, as a seller, would earn a total income of:  
\begin{equation}
    x_3(t+1) = f_{31}(t) \, x_1(t) + f_{32}(t) \, x_2(t) 
\end{equation}
over the time interval $[t,t+1]$, as a result of her selling other goods and services to Alice and Bob, $f_{31}(t)$ and $f_{32}(t)$ being the fractions of the buyers' respective incomes $x_1(t)$ and $x_2(t)$ that were paid to Carol. \\~\\
Using matrix algebra, the above three equations can be written as follows:
\begin{equation}
    \mathbf{x}(t+1) = \mathbf{F}_t \, \mathbf{x}(t) 
\end{equation}
where $\mathbf{x}(\tau) = [x_1(\tau), x_2(\tau), x_3(\tau)]^T$ is the income vector at time $\tau$, and the matrix:
\begin{equation}
    \mathbf{F}_{\tau} = [f_{ij}(\tau)]_{_{i,j\in \{1,2,3\}}} 
\end{equation}
is the $3 \times 3$ matrix whose entries are the above-mentioned fractions $f_{ij}(\tau)$ at time $\tau$. The superscript $\ ^T$ denotes, in the entire article, the transpose of a vector or a matrix. We should note that some of the entries in $\mathbf{F}_{\tau}$ might be equal to zero if there is no exchange between the corresponding agents. In particular, according to the above linear algebraic description of income circulation, the diagonal entries $f_{ii}(\tau)$ of the matrix $\mathbf{F}_{\tau}$ will be all zeroes, because agents don't sell things to themselves. But they can save money between one time instant and the next one. As business actors, they earn money by selling goods and services to others and, as living entities, they incur expenses to maintain themselves. The difference between the income earned and the expenses incurred would be the agents' savings. To be more specific, let us consider one individual agent $j$ which can be Alice, Bob or Carol. The total expenses that are incurred by $j$ between $t$ and $t+1$ can be calculated in the following way using the entries in the column $j$ of $\mathbf{F}_{\tau}$: 
\begin{equation}
    (\text{total expenses})_j = \sum_{\substack{i\in \{1,2,3\} \\ i \neq j}} f_{ij}(t) \, x_j(t) 
\end{equation}
Then, the difference:

\begin{equation}
\begin{split}
 x_j(t) - \, (\text{total expenses})_j & = x_j(t) \, - \sum_{\substack{i\in \{1,2,3\} \\ i \neq j}} f_{ij}(t) \, x_j(t)  \\
 & = \underbrace{\left( 1 - \sum_{\substack{i\in \{1,2,3\} \\ i \neq j}} f_{ij}(t) \right)}_{s_j(t)} \, x_j(t) \\
 \end{split}
\end{equation}
is what the agent $j$ would have saved over the time interval $[t,t+1]$.\\ 

\noindent Let us now redefine the diagonal entries $f_{jj}(\tau)$ for each agent $j \in \{1,2,3\}$ in the matrix $\mathbf{F}_{\tau}$ as: 

\[ f_{jj}(\tau) = s_j (\tau) = \left( 1 - \sum_{\substack{i\in \{1,2,3\} \\ i \neq j}} f_{ij}(\tau) \right)\]
and redefine the coordinates $x_i(\tau)$, $i\in \{1,2,3\}$, of the vector $\mathbf{x}(\tau)$ as the \textit{sum} of the income earned \textit{and} the savings made by agent $i$ between the time instants $\tau -1$ and $\tau$. With these new definitions of $f_{jj}(\tau)$ and $x_i(\tau)$, the matrix equation:  
\begin{equation}
    \mathbf{x}(t+1) = \mathbf{F}_t \, \mathbf{x}(t)  \label{Equ_8}
\end{equation}
remains valid with the understanding that the coordinates of $\mathbf{x}(\tau)$ at time $\tau$ represent the sum ``\textit{income} + \textit{savings}" made by the individual economic agents between $\tau-1$ and $\tau$, not just the \textit{income}. At this stage of building the linear algebraic model \eqref{Equ_8}, banks and financial institutions who would lend money to agents are not included. Because of that, the diagonal entries in the matrix $\mathbf{F}_{\tau}$ will remain non-negative. Note also that, with the new definitions of the diagonal entries and the coordinates of $\mathbf{x}$, the matrix $\mathbf{F}_{\tau}$ is column-stochastic.
\\~\\~\\

\noindent \textbf{\Large An agent-based income circulation model:}
\\

\noindent The gedankenexperiment we used in the above discussion involved three agents \,---\, Alice, Bob and Carol. But we can easily repeat this experiment and generalize it to $n$ agents with $n>3$, in which case the final version of income circulation model becomes:\\

\begin{equation}
    \mathbf{x}(t+1) = \mathbf{F}_t \, \mathbf{x}(t)  \label{equ_9}
\end{equation}
where $\mathbf{F}_{\tau}$ is a square $n \times n$ non-negative matrix that is column-stochastic, and that is referred to in this paper as the \textit{income circulation matrix} at time $\tau \geq 0$. It goes without saying that, for large values of $n$, the matrix $\mathbf{F}_{\tau}$ will likely have in it many entries that are zeroes, because it is very improbable in a large economy that every pair of agents engages in an act of selling and buying between the times $\tau$ and $\tau +1$.\\

\noindent Let us now specify the context within which the above matrix model \eqref{equ_9} would be valid:

\begin{enumerate}
    \item The economy at hand $\mathcal{E}$ is a closed economy made up of $n$ agents, and can be a small village, a region, or an entire country. The number $n$ can be in the hundreds, thousands, millions or billions of agents.
    \item An agent in this economy $\mathcal{E}$ can be:
    \begin{itemize}
        \item an individual who: (1) is an employee working for another agent (firm, individual, household, etc.), or (2) is running her own business (corner store, plumbing services, barber shop, handyman, and so on);
        \item a household made up of individuals who earn income, say the parents, and other ones who are dependent on the income earners;
        \item a firm which consists of a business or a collection of businesses that are, in principle, more sophisticated than the ones owned and operated by individuals. Those firms that are collections of businesses can be broken down into multiple agents in the matrix model we propose in this paper. For example, if the firm is a retail chain, then each retail store can be considered as a different agent in the income circulation matrix. Such a retail store sells goods and services to other agents in its neighborhood, while it buys goods and services from its corporate office which would be represented as a different agent in the income circulation matrix. 
    \end{itemize}
    
    \item At this initial stage of the development of the model, we assume that the agents in $\mathcal{E}$ cannot be banks or financial institutions who provide credit and, as a result, agent debt is not considered in this model. In general, the financial accounting concepts of \textit{liability} and \textit{asset} are intentionally omitted in the current model, but can be incorporated in future versions of the model. Similarly, no governments collecting taxes are included at this stage in the current model. 
    
    \item Time is discretized into uniform steps $t$, $t+1$, $t+2$, etc. There is no restriction on the length of these time steps which can be in the range of seconds, hours, days, or weeks, depending on the intensity of the economic activity in $\mathcal{E}$.
    \item For each integer $i \in \{1, 2, \cdots, n\}$, let $x_i(\tau) \geq 0$ be the sum of income earned and savings made by agent $i$ between times $\tau - 1$ and $\tau$. We will refer to $x_i(\tau)$ as the \textit{wealth} of agent $i$ at time $\tau$. The vector $\mathbf{x}(\tau) = [x_1(\tau), x_2(\tau), \cdots, x_n(\tau)]^T \in \Re_+^n$ is the wealth vector of agents in $\mathcal{E}$.
    
\end{enumerate}

\noindent Under the conditions specified in the above itemized list, the dynamics of the wealth vector  $\mathbf{x}$ are governed by the matrix equation \eqref{equ_9}. Using the rules of matrix algebra, we can show that the wealth vector $\mathbf{x}$ can be expressed as:
\begin{equation*}
    \mathbf{x}(t+1) = \left( \prod_{\tau=0}^{t} \mathbf{F}_{\tau} \right) \, \mathbf{x}(0)   
\end{equation*}
or, equivalently:\\
\begin{equation}
\boxed{\mathbf{x}(t) = \left( \prod_{\tau=0}^{t-1} \mathbf{F}_{\tau} \right) \, \mathbf{x}(0)}  \label{equa_10}
\end{equation}
where $\mathbf{x}(0)$ is the wealth vector of the economy $\mathcal{E}$ at an arbitrarily selected initial time instant $t=0$. The matrix equation \eqref{equa_10} constitutes what we believe to be a fundamental result about the dynamics of income circulation which, to our knowledge, has never been published in the economics literature, and which can be stated as follows:

\begin{center}
\fbox{\begin{minipage}{26em}
\vspace{0.4cm}
\textit{The dynamics of wealth distribution in a closed economy where no credit is available to the economic agents are governed by \textbf{in-homogeneous products of column-stochastic matrices}.}\\
\end{minipage}}
\end{center}

\noindent The sum $M=\sum_{i=1}^n x_i(0)$ of the coordinates of $\mathbf{x}(0)$ can be referred to as the \textit{monetary base}. Since the financial sector is  inexistent in $\mathcal{E}$, no money is created and, thus, $M$ remains constant over time. Using matrix algebraic rules, one can show that the sum of the coordinates of the wealth vector $\mathbf{x}(t)$ is invariant under transformations by in-homogeneous products of column-stochastic matrices. This ensures that the income circulation equation \eqref{equa_10} is consistent with the initial setup of $\mathcal{E}$. Also, in the cases where the length of the time steps are so short that no  business exchanges take place within a certain time interval $[\tau,\tau+1]$, the income circulation matrix $\mathbf{F}_{\tau}$ would just collapse to the identity matrix, as agents get to keep all their wealth to themselves during this time interval. Because the identity matrix is the neutral element for matrix multiplication, the matrix model \eqref{equa_10} is consistent with times steps being of any size. We also conjecture that one can derive from the matrix model \eqref{equa_10} many of the empirical observations that economists have highlighted about the issues of economic inequality, including the fact that wealth  distributions in societies follow, under some conditions on the structures of the income circulation matrices $\mathbf{F}_{\tau}$ ($\tau >0$), the Pareto law.\\
\\~\\

\noindent \textbf{\Large About the authors:}
\\

\noindent AG and JH are two Canadian systems scientists with background in mathematics, physical chemistry and engineering science. They live in Toronto, Ontario, and can be reached at the following e-mail address:

\begin{center} \textit{a2guerga@torontomu.ca}
\end{center}
Comments and feed-backs on this paper and its equations are welcomed and can be sent to the authors at the above e-mail address. \\

\printbibliography 

\end{document}